%% file: main.tex
\documentclass[runningheads]{llncs}
\input{macros}
\begin{document}
\title{Link Inference Attack on Privacy-Preserving Knowledge Graphs}
%
%
\author{ Emna Bouguerra\inst{1}\and
Ibtissam Harrouche\inst{1}\orcidlink{0009-0002-8249-3672}\and
Ferran Alborch\inst{1}\orcidlink{0000-0002-3563-9133}\and
Melek Önen\inst{1}\orcidlink{0000-0003-0269-9495}}
\authorrunning{E. Bouguerra et al.}
%
\institute{
EURECOM, Sophia Antipolis, France\\[0.5em]
\email{\{emna.bouguerra,ibtissam.harrouche,ferran.alborch,melek.onen\}@eurecom.fr}\
}
\maketitle              
\begin{abstract}
Knowledge Graphs (KGs) are widely used to store and share structured
information across sensitive domains such as healthcare, finance, and
social networks. A common privacy practice is to delete sensitive
relations before publishing the graph, under the assumption that
removing edges is sufficient to prevent their recovery. In this paper,
we challenge this assumption and show that even when a relation is
fully or partially hidden, its existence leaves structural traces in
the public graph that can be exploited to recover it with high
accuracy. To this end, we propose a link inference attack that
operates on the topology of the public graph, and evaluate it under
two privacy scenarios that differ in how the adversary
exploits the knowledge available to him. In the first 
setting where the adversary exploits all topological information, the attack achieves near-perfect discrimination (AP = 0.949,
ROC-AUC = 0.999), while in the more realistic one where the adversary makes use of some semantic information, it recovers up to 74\%
of hidden edges. Building on these results, we further
conduct a structural analysis to identify which topological properties
of the graph drive the attack success, revealing that privacy risk is
not uniform across entities and that certain structural patterns make
specific relations significantly more vulnerable to inference than
others.

\end{abstract}
\input{Sections/intro}
\input{Sections/background}

\input{Sections/related_work}

\input{Sections/new_attacks}
\input{Sections/Structural_Leakage_Analysis} 
\input{Sections/conclusion}

\subsubsection{\ackname} This work has been conducted within the project TRUMAN. The research leading to these results has received funding from HORIZON-CL4-2024-HUMAN-03, under Grant Agreement no 101214000.

%
%
%
\bibliographystyle{splncs04}
\bibliography{mybibliography}

\appendix
\section{Evaluation Metrics}
\label{app:metrics}

\paragraph{Precision and Recall.}
Given a set of retrieved instances, precision measures the fraction
of retrieved positive instances that are true positives:
\begin{equation}
    P = \frac{|\text{TP}|}{|\text{TP}| + |\text{FP}|},
\end{equation}
Recall measures the fraction of true positives among true labels:
\begin{equation}
    R = \frac{|\text{TP}|}{|\text{TP}| + |\text{FN}|},
\end{equation}
where TP, FP, and FN denote true positives, false positives,
and false negatives respectively.

\paragraph{Average Precision (AP).}
The area under the precision-recall curve, summarizing model
performance across all classification thresholds:
\begin{equation}
    \text{AP} = \sum_{k=1}^{K} P(k) \cdot \Delta R(k),
\end{equation}
where $P(k)$ and $R(k)$ are the precision and recall at the
$k$-th threshold, and $\Delta R(k) = R(k) - R(k-1)$.

\paragraph{ROC-AUC.}
The area under the receiver operating characteristic curve,
measuring the probability that a randomly chosen positive instance
is ranked higher than a randomly chosen negative one:
\begin{equation}
    \text{ROC-AUC} = \frac{\sum_{x^+ \in \mathcal{P}}
    \sum_{x^- \in \mathcal{N}}
    \mathbf{1}[s(x^+) > s(x^-)]}
    {|\mathcal{P}| \cdot |\mathcal{N}|},
\end{equation}
where $\mathcal{P}$ and $\mathcal{N}$ are the sets of positive
and negative instances, and $s(x)$ is the predicted score
for instance $x$.

\paragraph{Rank.}
Given a query $q$ and a list of candidates ranked by descending
predicted score, $\text{rank}(q)$ denotes the position of the
first true positive in the ranked list. A rank of 1 means the
true positive is the top-ranked candidate.

\paragraph{Hits@$k$.}
The fraction of queries for which there is a true positive candidate ranked within the first $k$ positions
\begin{equation}
    \text{Hits@}k = \frac{1}{|\mathcal{Q}|}
    \sum_{q \in \mathcal{Q}}
    \mathbf{1}\!\left[\text{rank}(q) \leq k\right],
\end{equation}
where $\mathcal{Q}$ is the set of queries.

\paragraph{Mean Reciprocal Rank (MRR).}
The mean of the reciprocal rank of the first true positive
across all queries:
\begin{equation}
    \text{MRR} = \frac{1}{|\mathcal{Q}|}
    \sum_{q \in \mathcal{Q}}
    \frac{1}{\text{rank}(q)}.
\end{equation}
A higher MRR indicates that true positives are consistently
ranked near the top.
\end{document}

%% file: macros.tex
\usepackage[T1]{fontenc}
\usepackage{graphicx}
\usepackage{amsmath,amsfonts,amscd}
\usepackage{bm}
\usepackage[dvipsnames]{xcolor}
\usepackage{graphicx}
\usepackage{mathtools}
\usepackage{pifont}
\usepackage{verbatim}
\usepackage{multicol,multirow}
\usepackage{orcidlink}

\usepackage{algorithm}
\usepackage{algorithmic}

\usepackage{tkz-graph}
\usetikzlibrary{shapes.symbols}
\usetikzlibrary{arrows.meta}
\usetikzlibrary{positioning}
\usetikzlibrary{patterns}

\newcommand{\melek}[1]{\textcolor{violet}{\fbox{M\"O}#1}}
\newcommand{\IH}[1]{\textcolor{cyan}{\fbox{IH}#1}}


%% file: Sections/intro.tex
\section{Introduction} \label{sec:intro}

Knowledge Graphs (KGs) are, nowadays, widely used to store and share
structured information across various domains including healthcare \cite{xang2019knowledge}, fi-
nance \cite{ji2021survey}, and social networks \cite{ji2021survey}. Their widespread adoption unfortunately begins to raise privacy issues: Indeed some part of the knowledge graph may include privacy-sensitive or confidential information:  these could include relations such as a person's medical condition or his/her salary. 

Stakeholders who own the graph, naturally wish to make use of the knowledge to train accurate machine learning models without disclosing the privacy-sensitive links/relations. A recent privacy protection practice \cite{10198395} consists of publishing the knowledge graph and further making use of it, once all sensitive edges are deleted. 

In this paper, we study the consequences of such a naive protection and further show that even when a sensitive relation is deleted, the remaining information and the structure of the graph can easily help an adversary to recover it. Indeed, we design a link inference attack that by processing the information on the topology of the graph and sometimes additionally some embedding information aims at inferring the hidden links of the KG. More specifically, we propose two attack scenarios that
differ in how the adversary exploits the knowledge available to it: while the first one is oblivious to the semantics of the graph and makes use of all the triples in the graph, the second one is more optimized since it only processes links with entities already connected with sensitive relations.  

We implement and evaluate these two scenarios and show that they result in significant leakage with non significant cost. Building on these observations, we further conduct a structural analysis to identify which topological properties of the graph drive the success of the attack. We try to extract these using Principal Component Analysis (PCA). We further observe that the leakage is not uniform. For example. highly connected entities in the graph are more exposed to these attacks.

 \paragraph{Paper outline.} Section \ref{sec:background} provides background information on knowledge graphs. Related work is discussed in section \ref{sec:relatedwork}. The two proposed attacks are described and evaluated in section \ref{sec:New Attacks}. Section \ref{sec:Structural Leakage Analysis} extends the study to identify the actual structural properties that influence privacy leakage and study this leakage across different topological patterns. Finally, conclusive remarks are provided in section \ref{sec:conclusion}. 



%% file: Sections/background.tex
\section{Preliminaries} \label{sec:background}
  \subsection{Knowledge Graphs}

  A \textbf{Knowledge Graph} (KG) is defined as a set of triples $\mathcal{G} = \{(h, r, t)\in\mathcal{T}\coloneq\mathcal{E}\times\mathcal{R}\times\mathcal{E}\}$, where $\mathcal{E}$ is a set of entities and $\mathcal{R}$ is a set of relations. We denote as $h\in\mathcal{E}$ the \emph{head} entity, as $t \in \mathcal{E}$ the \emph{tail} entity, and as $r\in\mathcal{R}$ the \emph{relation} type connecting them. Each triple $(h, r, t)$ represents a directed {edge} from $h$ to $t$ labeled with relation $r$. KGs are a useful way to encode structured data, and as such have been used for a plethora of machine learning applications, such as recommendation systems \cite{wang2019KGAT}, search engines \cite{10393039}, question answering \cite{9960856}, and clinical decision support \cite{xang2019knowledge}.

\subsection{Knowledge graph structural features} \label{sec:structure}

Knowledge graphs can be characterized by several entity-related structural metrics that capture the role and position of entities within the graph topology. In the following, we introduce the key metrics used in our analysis.

\subsubsection{Node Degree}
The degree of an entity $e \in \mathcal{E}$ quantifies its connectivity within the graph, i.e. the amount of other entities involved in a triple with $e$. We distinguish three types of degree:
\begin{itemize}
    \item In-degree $\text{deg}_{in}(e) = |\{(h, r, e) \in \mathcal{T}\}|$: denotes the number of entities involved in a triple where $e$ is a tail.
    \item Out-degree $\text{deg}_{out}(e) = |\{(e, r, t) \in \mathcal{T}\}|$: denotes the number of entities involved in a triple where $e$ is a head.
    \item Total degree $\text{deg}(e) = \text{deg}_{in}(e) + \text{deg}_{out}(e)$: the total number of entities involved in a triple with $e$.
\end{itemize}
Entities with higher degree values tend to be more central and well-connected within the graph structure.

\subsubsection{Incident Relation Counts}
Beyond overall connectivity, we can examine how frequently a specific entity $e$ participates in a particular relation type $r$. We define:
\begin{itemize}
    \item $\text{count}_r^{head}(e) = |\{(e, r, t) \in \mathcal{T}\}|$: the number of triples where $e$ appears as the head with relation $r$.
    \item $\text{count}_r^{tail}(e) = |\{(h, r, e) \in \mathcal{T}\}|$: the number of triples where $e$ appears as the tail with relation $r$
\end{itemize}
These counts provide a more fine-grained view of an entity's involvement in specific relationship types.

\subsubsection{Jaccard Similarity}
To compare entities structurally, we use the Jaccard similarity coefficient, which measures the overlap between their neighborhoods. Given two entities $e_1, e_2$, we first define the neighborhood of an entity $e$ as:
\begin{equation}
N(e) = \{v \in \mathcal{E} \mid (e, r, v) \in \mathcal{G} \text{ or } (v, r, e) \in \mathcal{G}\}.
\end{equation}
$|S|$ denotes the cardinality of a given set $S$.
The Jaccard similarity between $e_1$ and $e_2$ is then:
\begin{equation}
J(e_1, e_2) = \frac{|N(e_1) \cap N(e_2)|}{|N(e_1) \cup N(e_2)|}.
\end{equation}

This coefficient ranges from 0 (indicating no shared neighbors) to 1 (indicating identical neighborhoods). A higher Jaccard similarity suggests that two entities occupy similar structural positions within the graph.

\subsection{Knowledge Graph Embeddings}
Knowledge Graph Embedding (KGE) models learn vector representations of
entities and relations in $\mathbb{R}^d$, capturing the structural and
semantic regularities of the graph, with applications in link
prediction~\cite{bordes2013translating}, question
answering~\cite{9960856}, and recommendation
systems~\cite{wang2019KGAT}. Among these, TransE~\cite{bordes2013translating}
represents each relation as a translation vector, optimizing:
\begin{equation}
    f(h, r, t) = \|\mathbf{v}_h + \mathbf{v}_r - \mathbf{v}_t\|,
\end{equation}
where $\mathbf{v}_h, \mathbf{v}_r, \mathbf{v}_t \in \mathbb{R}^d$ are
the learned embeddings of the head entity, relation, and tail entity
respectively. A low score indicates a likely true triple. TransE produces
meaningful relation embeddings in which semantically similar relations
cluster together, a property we exploit in Scenario 2
(Section~\ref{sec:New Attacks}).

%% file: Sections/related_work.tex
\section{Related Work} \label{sec:relatedwork}

Knowledge graphs have emerged as a fundamental means to
represent structured knowledge across diverse domains, from biomedical
research to recommendation systems. As their deployment grows, so does
the concern do: releasing even a partial view of a KG may expose
sensitive information that was never intended to be public. This section
reviews the most relevant work on inference attacks against
graph-structured data and KGs specifically.

\paragraph{Inference Attacks on Graph-Structured Data.}
A foundational result by Backstrom et al.~\cite{backstrom2007wherefore}
demonstrates that even an anonymized social network leaks edge-level
information through structural patterns, alone: given only the topology
of an anonymized graph, an adversary can determine whether specific
edges exist between targeted pairs of nodes. This establishes a key
principle that our work extends to Knowledge Graphs: structural
information alone is sufficient to mount a privacy attack, without any
model access or semantic knowledge. Duddu et al.~\cite{duddu2020quantifying}
are among the first to systematically quantify privacy leakage in graph
embeddings, showing that GNN(Graph Neural Network) outputs expose sensitive node-level
information through membership inference attacks. He et
al.~\cite{he2021stealing} propose the first link stealing attacks
against GNN models, demonstrating that black-box access to a trained
model suffices to infer whether specific edges exist in the training
graph. Wu et al.~\cite{wu2022linkteller} further show that private
edges can be recovered from GNNs via influence analysis, even under
differentially private training mechanisms.

\paragraph{Inference Attacks on Knowledge Graphs.}
Wang et al.~\cite{wang2021membership} conduct the first empirical study
of membership inference attacks against KGE models, showing that these
models leak information about their training triples across multiple
benchmark datasets including sensitive medical and financial KGs.
Hu et al.~\cite{hu2023quantifying} extend this line of work to the
federated setting, proposing inference attacks against federated KGE
models and demonstrating that collaborative training does not eliminate
privacy risks. Our work is complementary to these efforts but targets
a fundamentally different attack surface: rather than exploiting a
trained model, we show that the raw structure of the public graph alone
is sufficient to recover suppressed relations with high accuracy,
without any model access or embedding.

While our attack builds on structural features also used in link prediction, the two tasks differ fundamentally. Link prediction aims to complete an incomplete graph, whereas we target edges that were deliberately suppressed for privacy. Our attacker operates with limited supervision from $\mathcal{T}_{known}$, requires no access to a trained model unlike ~\cite{wu2022linkteller} ~\cite{he2021stealing} and in Scenario 2 exploits relation-type semantics via KGE embeddings, a property specific to KGs that has no equivalent in homogeneous graphs.

%% file: Sections/new_attacks.tex
\section{Link Inference Attack on KGs} \label{sec:New Attacks}
In this section, we present a link inference attack designed to recover sensitive relations that have been deliberately hidden from a Knowledge Graph. The adversary exploits the publicly observable structure of the graph, only.  

\subsection{Threat Model}

We model an adversary who seeks to recover a sensitive relation $r^{\star}$
that has been suppressed from a publicly released Knowledge Graph
$\mathcal{G}_{pub}$. The attacker has two sources of information: (1) the
full structure of $\mathcal{G}_{pub}$, including all non-sensitive relations,
and (2) a partial set of known $r^{\star}$ triples representing facts that
are already publicly available for a subset of entities, whether through
prior releases, incomplete anonymization, or other disclosures. The attacker's goal is to exploit these two sources
to infer the remaining hidden triples of $r^{\star}$.

Formally, let $\mathcal{T}^{\star}\coloneq\{(h,r^{\star},t)\in\mathcal{T}\}$ denote the set of triples with sensitive relation $r^{\star}$, and $\mathcal{T}_{known}\subseteq\mathcal{T}^{\star}$ the subset known by the adversary. The attacker then observes:
\begin{equation}
    \mathcal{G}_{pub} = \{(h,r,t) \in \mathcal{T} \mid r \neq r^{\star}\}
    \;\cup\; \mathcal{T}_{known},
\end{equation}
and aims to recover $\mathcal{T}^{\star} \setminus \mathcal{T}_{known}$. The attacker operates solely on anonymous entity identifiers, with no access to real-world entity labels or semantic knowledge.

The two scenarios we consider differ only in \textbf{how the adversary constructs the candidate space} from $\mathcal{G}_{pub}$, as the supervised training signal is the same in both cases.

\subsection{Attack Description}\label{sec:Description}

\begin{table}[t]
\centering
\caption{Comparison of the two attack scenarios.}
\label{tab:scenario_comparison}
\begin{tabular}{lclcl}
\hline
\textbf{Property} & & \textbf{Scenario 1} & & \textbf{Scenario 2} \\
\hline
Candidate space   & & All co-occurring pairs & & Semantically guided expansion \\
Strategy          & & Brute-force            & & Proxy-guided \\
Attack’s cost       & & Expensive              & & Optimized \\
Semantic filtering & & None                  & & Via relation similarity \\
\hline
\end{tabular}
\end{table}

The attack exploits the following property of KGs: even when
$r^{\star}$ is removed, its existence leaves structural traces in
$\mathcal{G}_{pub}$. Entities connected by $r^{\star}$ tend to share
similar structural features: they share common neighbors, co-appear
in related relations, and exhibit comparable connectivity patterns.
These regularities are consistent enough to be learned from the known
triples $\mathcal{T}_{known}$ and generalized to unseen pairs.

Concretely, the adversary uses $\mathcal{T}_{known}$ as labeled examples to train a binary classifier that distinguishes $r^{\star}$ edges from non-edges, based solely on structural features extracted from $\mathcal{G}_{pub}$ (see \ref{sec:structure}). The trained classifier is then applied to a set of candidate pairs to rank the most likely hidden edges.

The attack follows four steps: it (1) constructs a candidate set
$\mathcal{P}$ of entity pairs, (2) extracts $\mathcal{P}$'s structural features from $\mathcal{G}_{pub}$, (3) trains a binary classifier
where positive examples are the known triples $\mathcal{T}_{known}$
and negative examples are all remaining pairs in $\mathcal{P}$, and (4) ranks candidates by predicted score. The two scenarios differ in Step~1, leading to distinct strategies, resulting in different costs and including semantic filtering or not as summarized in Table~\ref{tab:scenario_comparison}.

\subsection{The two attack scenarios}
\label{sec:impl}
\subsubsection{Scenario 1: Na\"ive Attack }

The adversary constructs the candidate space without any semantic filtering:
every entity pair co-occurring in at least one triple in $\mathcal{G}_{pub}$
is considered a potential candidate:
\begin{equation}
    \mathcal{P} = \{(h, t) \mid \exists\, r : (h, r, t) \in \mathcal{G}_{pub}\}.
\end{equation}
This yields an exhaustive but very large candidate space, in which true
$r^{\star}$ edges typically represent only a small fraction of all candidates,
reflecting the natural sparsity of Knowledge Graphs.
Each pair is labeled against the original graph prior to masking:
\begin{equation}
    y(h,t) =
    \begin{cases}
        1 & \text{if } (h, r^{\star}, t) \in \mathcal{G}, \\
        0 & \text{otherwise}.
    \end{cases}
\end{equation}
$\mathcal{P}$ is split into train, validation, and test sets via stratified
sampling. Structural features are extracted from $\mathcal{G}_{pub}$ for
each pair (Section~\ref{sec:structure}). We train and compare Logistic
Regression and LightGBM, with hyperparameters tuned on the validation set.
Since this scenario reduces to a binary classification task over existing
pairs in $\mathcal{G}_{pub}$, we report AP and ROC-AUC as evaluation metrics.

\subsubsection{Scenario 2: Semantic-Aware Attack}

In this scenario, the adversary leverages $\mathcal{T}_{known}$, the set
of $r^{\star}$ triples already known, as a seed to build a focused and
semantically coherent candidate space, rather than enumerating all pairs
blindly. Visible edges serve directly as training signal; the key challenge
is constructing a focused candidate space for the private heads.

\paragraph{Relation Similarity via KGE.}
A TransE model~\cite{bordes2013translating} is trained on $\mathcal{G}_{pub}$ to rank relations by
proximity to $r^{\star}$. For each relation $r$, let $\mathbf{v}_r \in
\mathbb{R}^d$ denote its learned embedding. The cosine similarity between
$\mathbf{v}_r$ and the target relation embedding $\mathbf{v}_{r^{\star}}$ is:

\begin{equation}
    \text{sim}(r, r^{\star}) =
    \frac{\mathbf{v}_r \cdot \mathbf{v}_{r^{\star}}}
    {\|\mathbf{v}_r\| \cdot \|\mathbf{v}_{r^{\star}}\|}.
\end{equation}
The top-$K$ most similar relations form the proxy set $\mathcal{R}_{sim}$.

\paragraph{Candidate Set Expansion.}
Starting from seed sets derived from visible edges,
\begin{equation}
    \mathcal{F}_{seed} = \{h \mid (h,r^{\star},t) \in \mathcal{T}_{known}\},
    \quad
    \mathcal{G}_{seed} = \{t \mid (h,r^{\star},t) \in \mathcal{T}_{known}\}
\end{equation}
let $\mathcal{H}_r$ and $\mathcal{T}_r$ denote the head and tail entities
of relation $r$ in $\mathcal{G}_{pub}$. Each proxy relation $r \in
\mathcal{R}_{sim}$ is classified as \emph{head-like} or \emph{tail-like}
based on its overlap with the seed sets: 
\begin{equation}
    \mathcal{R}_{head\text{-}like} = \left\{ r \in \mathcal{R}_{sim} \;\Big|\;
    \frac{|\mathcal{H}_r \cap \mathcal{F}_{seed}|}{|\mathcal{H}_r|}
    \geq \alpha \;\text{ and }\;
    |\mathcal{H}_r \cap \mathcal{F}_{seed}| \geq n_{h}
    \right\},
\end{equation}
\begin{equation}
    \mathcal{R}_{tail\text{-}like} = \left\{ r \in \mathcal{R}_{sim} \;\Big|\;
    \frac{|\mathcal{T}_r \cap \mathcal{G}_{seed}|}{|\mathcal{T}_r|}
    \geq \beta \;\text{ and }\;
    |\mathcal{T}_r \cap \mathcal{G}_{seed}| \geq n_{t}
    \right\}.
\end{equation}
The fraction thresholds $\alpha, \beta$ ensure that a proxy relation 
involves entities of the same semantic type as $r^{\star}$. The minimum 
count thresholds $n_h, n_t$ prevent relations with very few participants 
from qualifying, as a small relation can achieve a high overlap fraction 
purely by chance rather than by genuine similarity. Both conditions must 
hold simultaneously for a relation to qualify. The enriched candidate 
sets are then:
\begin{equation}
    \tilde{\mathcal{F}} = \mathcal{F}_{seed} \cup
    \bigcup_{r \in \mathcal{R}_{head\text{-}like}} \mathcal{H}_r, \qquad
    \tilde{\mathcal{G}} = \mathcal{G}_{seed} \cup
    \bigcup_{r \in \mathcal{R}_{tail\text{-}like}} \mathcal{T}_r.
\end{equation}
The attacker then infers that any entity in $\tilde{\mathcal{F}}$ with no
visible $r^{\star}$ edges is a private head:
$\mathcal{F}_{attack} = \tilde{\mathcal{F}} \setminus \mathcal{F}_{seed}$.

\paragraph{Candidate Pair Construction.}
Two pair sets are constructed:
\begin{itemize}
    \item \textbf{Training/validation} ($\mathcal{P}_{vis}$):
    $\mathcal{F}_{seed} \times \tilde{\mathcal{G}}$, split 80/20 with
    stratified sampling. Labels are assigned from the full unmasked graph
    $\mathcal{G}$ as ground truth: $y(h,t) = 1$ if $(h, r^{\star}, t) \in
    \mathcal{G}$, and $0$ otherwise.
    \item \textbf{Test} ($\mathcal{P}_{attack}$):
    $\mathcal{F}_{attack} \times \tilde{\mathcal{G}}$. Labels are unknown
    to the attacker and used only for offline evaluation.
\end{itemize}

\paragraph{Classifier Training and Ranking.}
The same five structural features as Scenario~1 are extracted from
$\mathcal{G}_{pub}$ for each candidate pair. A LightGBM classifier is
trained on $\mathcal{P}_{vis}$ with hyperparameters tuned on the validation
split. For each private head $h \in \mathcal{F}_{attack}$, candidate tails
in $\tilde{\mathcal{G}}$ are ranked by descending score. Since this scenario
is framed as a ranking task over semantically guided candidate pairs, we
primarily report Hits@1 and MRR, which directly measure ranking quality.
AP is also reported for completeness, though it is less informative in
ranking settings where the candidate space is focused and semantically
coherent.

\subsection{Experimental study}

\paragraph{\textbf{Dataset.}}
We conduct our experiments on FB15k-237~\cite{10.1145/1376616.1376746},
a widely used Knowledge Graph benchmark derived from Freebase. It
contains 14,541 entities, 237 relation types, and 310,116 triples,
covering diverse domains including film, sports, music, and people.
We target three relations of varying frequency:
\texttt{/film/film/} \texttt{genre} (medium frequency),
\texttt{/award/.../award\_nominee} (high frequency),\\[0.1em] and
\texttt{/people/.../specialization\_of} (low frequency).
\paragraph{\textbf{Evaluation Metrics.}}

We evaluate our attack using four metrics: Average Precision (AP),
ROC-AUC, Hits@1, and Mean Reciprocal Rank (MRR), whose formal
definitions are provided in Appendix~\ref{app:metrics}. AP and
ROC-AUC are used for Scenario~1, framed as a binary classification
task. Hits@1 and MRR are primarily used for Scenario~2, framed as
a ranking task, with AP also reported for completeness.

\paragraph{\textbf{Scenario 1:Na\"ive Attack.}}

Table~\ref{tab:attack1_results} reports the performance of the two classifiers
against a random baseline on \texttt{/film/film/genre} (FB15k-237).

\begin{table}[t]
\centering
\caption{Na\"ive Attack results on \texttt{/film/film/genre} (FB15k-237).}
\begin{tabular}{lccc}
\hline
\textbf{Model} & \hspace{3pt}\textbf{Test AP}\hspace{3pt} & \hspace{3pt}\textbf{ROC-AUC}\hspace{3pt} &
\hspace{3pt}\textbf{Gain over Random}\hspace{3pt} \\
\hline
Random Baseline     & 0.030 & 0.500 & ---          \\
Logistic Regression & 0.438 & 0.958 & $\times$14.6 \\
LightGBM            & 0.949 & 0.999 & $\times$31.6 \\
\hline
\end{tabular}
\label{tab:attack1_results}
\end{table}

Both classifiers substantially outperform the random baseline. LightGBM
achieves near-perfect discrimination (AP = 0.949, ROC-AUC = 0.999,
$\times$31.6 over random), while Logistic Regression already yields strong
performance (AP = 0.438, ROC-AUC = 0.958), confirming that even a linear
model can exploit structural leakage. The gap between the two models suggests
that the relationship between structural features and hidden link existence is
non-linear, which the tree-based model captures more effectively. These
results hold under severe class imbalance ($\sim$3\% positives), as reflected
by the high ROC-AUC across both models.

The main limitation of this scenario is its cost: on FB15k-237, the candidate
space reaches 283,831 pairs for a single relation, many of which are pairs
that are clearly unrelated to $r^{\star}$ and thus trivially classified as
negative. This motivates the more targeted approach of the attack in
Scenario~2.

\paragraph{\textbf{Scenario 2: Semantic-Aware Attack.}}
Table~\ref{tab:coverage} reports the coverage of $\tilde{\mathcal{F}}$ and
$\tilde{\mathcal{G}}$ across three target relations. For
\texttt{/film/film/genre}, expansion recovers 99.8\% of all true head
entities, up from 70.0\% for the seed alone.

\begin{table}[t]
\centering
\caption{Head and tail coverage of the expanded candidate sets across three
target relations. Seed$\to$All: fraction of true entities recovered by the
seed set alone. Exp.$\to$All: fraction recovered after expansion.}
\label{tab:coverage}
\begin{tabular}{lcccc}
\hline
\multirow{2}{*}{\textbf{Target Relation}}
    & \multicolumn{2}{c}{\textbf{Head Coverage}}
    & \multicolumn{2}{c}{\textbf{Tail Coverage}} \\
    & Seed/All & Exp./All & Seed/All & Exp./All \\
\hline
\texttt{.../specialization\_of} & 69.7\% & 73.1\% & 85.7\% & 97.1\% \\
\texttt{.../award\_nominee}     & 70.0\% & 74.8\% & 92.9\% & 98.9\% \\
\texttt{/film/film/genre}       & 70.0\% & 99.8\% & 98.4\% & 99.2\% \\
\hline
\end{tabular}
\end{table}
Table~\ref{tab:attack2_results} reports performance across three target
relations of varying frequency in FB15k-237.

\begin{table}[t]
\centering
\caption{Semantic-aware Attack results across three target relations in
FB15k-237. Heads: number of private heads with at least one true hidden edge
in the test set.}
\begin{tabular}{llccccc}
\hline
\textbf{Relation} & \hspace{3pt}\textbf{Freq.}\hspace{3pt} & \hspace{3pt}\textbf{AP}\hspace{3pt} & \hspace{3pt}\textbf{MRR}\hspace{3pt}
& \hspace{3pt}\textbf{Hits@1}\hspace{3pt} & \hspace{3pt}\textbf{Heads}\hspace{3pt} & \hspace{3pt}\textbf{Gain}\hspace{3pt} \\
\hline
\texttt{.../award\_nominee}      & High   & 0.3241 & 0.5597 & 45.54\%           & 112 & $\times$648.2 \\
\texttt{/film/film/genre}        & Med.   & 0.2935 & 0.8274 & 74.07\%           & 563 & $\times$13.3  \\
\texttt{.../specialization\_of}  & Low    & 0.0005 & 0.0659 & \phantom{0}0.00\% & 3   & $\times$21.7  \\
\hline
\end{tabular} 

\label{tab:attack2_results}
\end{table}

Results reveal a clear dependence on relation frequency. For
\texttt{/film/film/} \texttt{genre} (medium frequency), the attack achieves 74.07\%
Hits@1 and MRR = 0.8274: for nearly three quarters of hidden films, the true
genre is ranked first. For \texttt{/award/.../award\_nominee} (high
frequency), Hits@1 is lower (45.54\%) but the gain over random is the
highest ($\times$648.2), driven by the large tail space. For
\texttt{/people/.../specialization\_of} (low frequency), the attack largely
fails (0\% Hits@1, MRR = 0.066): with only 122 triples in the full KG and
3 evaluable private heads, structural features provide insufficient signal.

The lower performance of Scenario 2 relative to Scenario 1 is expected by design. Scenario 1 trains on all co-occurring pairs in $\mathcal{G}_{pub}$ with full stratified supervision, while Scenario 2 trains only on visible heads and must rank over a candidate space that it partially reconstructs from semantic proximity. The performance gap therefore reflects the cost of a more realistic threat model, where the attacker has no prior knowledge of the full candidate space.

%% file: Sections/Structural_Leakage_Analysis.tex
\section{Structural Leakage Analysis} \label{sec:Structural Leakage Analysis}
   In this section, we conduct two complementary studies to better understand the origin
  of the leakage. First, we apply PCA to the five structural features to examine
  their information content and inter-dependencies. Second, we show that privacy risk is
  not uniformly distributed across entities, and that certain structural properties make
  some entities significantly more vulnerable than others.

  \subsection{PCA on Structural Features}
Our attack relies on five structural features computed for each candidate
pair $(h, t)$: \texttt{deg\_head}, \texttt{deg\_tail},
\texttt{inc\_rel\_head}, \texttt{inc\_rel\_tail}, and Jaccard similarity
between their one-hop neighborhoods. To understand the information content
of these features and their interdependencies, we apply Principal Component
Analysis (PCA) to the candidate pairs of Scenario~1

\begin{figure}[t]
      \centering
      \includegraphics[width=1\textwidth]{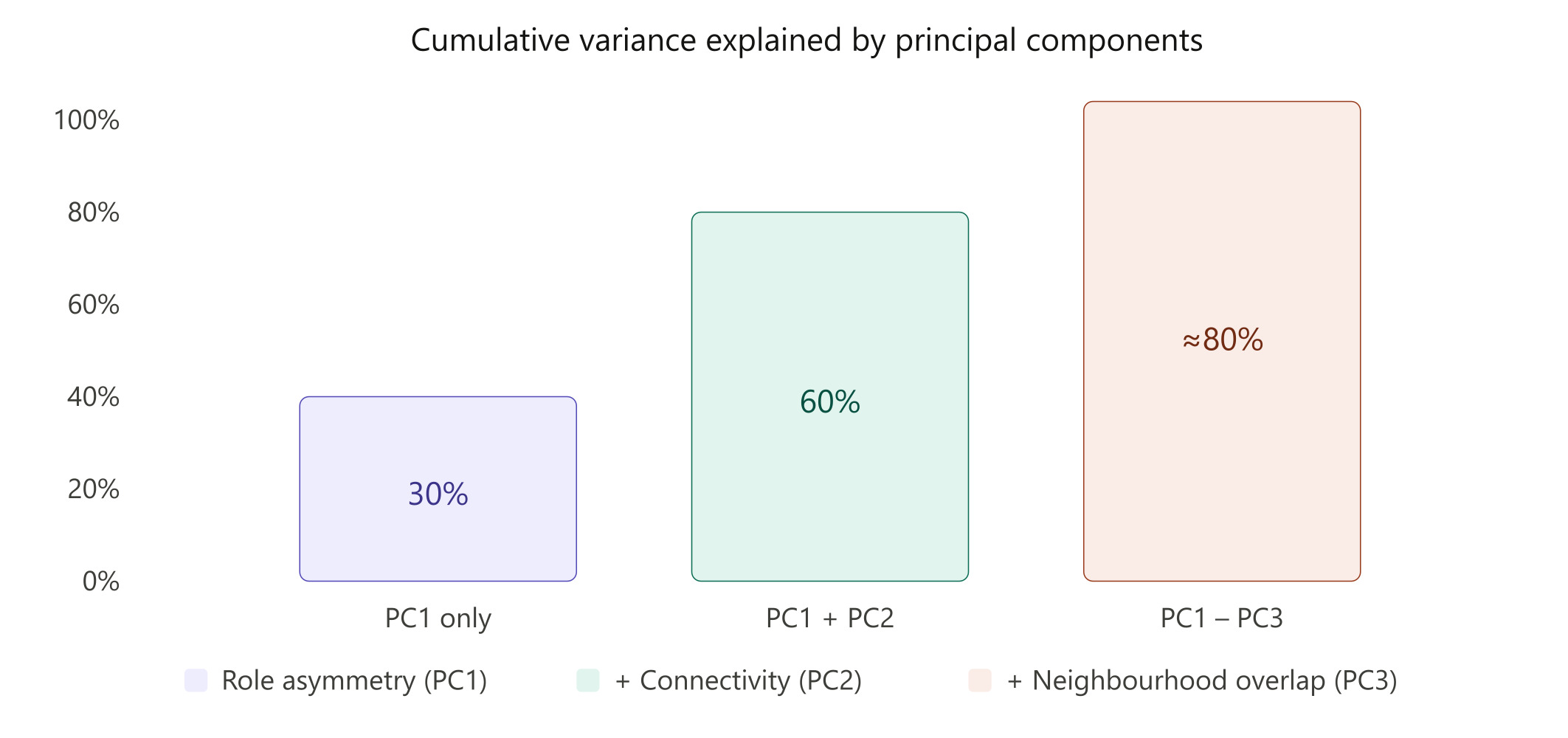}
      \caption{Cumulative  variance of the first three principal components.
      Three components capture approximately 80\% of the total variance in the five
      structural features.}
      \label{fig:pca_variance}
  \end{figure}

    As shown in Figure~\ref{fig:pca_variance}, the first three principal components capture
  approximately 80\% of the total variance, confirming that the five structural features
  are correlated and their information content compresses into three latent dimensions.

The loadings for all three components are summarized in Table~\ref{tab:pca_loadings}. 
\newline
  \textbf{PC1 — Head/Tail Asymmetry.} The first component loads positively
on \texttt{deg\_head} ($+0.573$) and \texttt{inc\_rel\_head} ($+0.557$),
and negatively on \texttt{deg\_tail} ($-0.426$) and
\texttt{inc\_rel\_tail} ($-0.402$), capturing the connectivity imbalance
between the two endpoints: high scores indicate a hub head paired with a
peripheral tail, and vice versa.
\newline
\textbf{PC2 — Overall Connectivity.} The second component loads positively
on all degree and incident relation features, with a small negative loading
on Jaccard ($-0.282$). Unlike PC1, it does not distinguish between head and tail it simply captures whether both entities are well-connected.
\newline
 \textbf{PC3 — neighborhood Overlap.} The third component is dominated
by Jaccard similarity ($+0.926$), capturing how much the local
neighborhoods of the head and tail overlap, independently of their
overall connectivity.

  \begin{table}[t]
  \centering
  \caption{Feature loadings for the first three principal components.}
  \label{tab:pca_loadings}
  \begin{tabular}{lccc}
  \hline
  \textbf{Feature} & \hspace{6pt}\textbf{PC1}\hspace{6pt} & \hspace{6pt}\textbf{PC2}\hspace{6pt} & \hspace{6pt}\textbf{PC3}\hspace{6pt} \\
  \hline
  \texttt{deg\_head}           & $+0.573$ & $+0.416$ & $-0.040$ \\
  \texttt{incident\_rel\_head} & $+0.557$ & $+0.430$ & $+0.129$ \\
  \texttt{deg\_tail}           & $-0.426$ & $+0.547$ & $+0.085$ \\
  \texttt{incident\_rel\_tail} & $-0.402$ & $+0.514$ & $+0.341$ \\
  \texttt{jaccard}             & $+0.135$ & $-0.282$ & $+0.926$ \\
  \hline
  \end{tabular}
  \end{table}

  These three components show that structural leakage operates along three independent axes:
  the \textit{imbalance} between head and tail connectivity, the \textit{overall connectivity}
  of both endpoints, and the \textit{neighborhood overlap} between them.

   \subsection{Privacy Risk Across Entities}
The attack does not affect all entities equally. A node that is highly
connected in the graph is structurally more exposed than one with few
connections, as its dense neighborhood provides the attacker with stronger
signal. For our target relation, this translates directly to genre
popularity: a film whose genres are shared by many other films is easier
to target than one whose genres are rare.

\begin{figure}[t]
      \centering
      \includegraphics[width=0.9\textwidth]{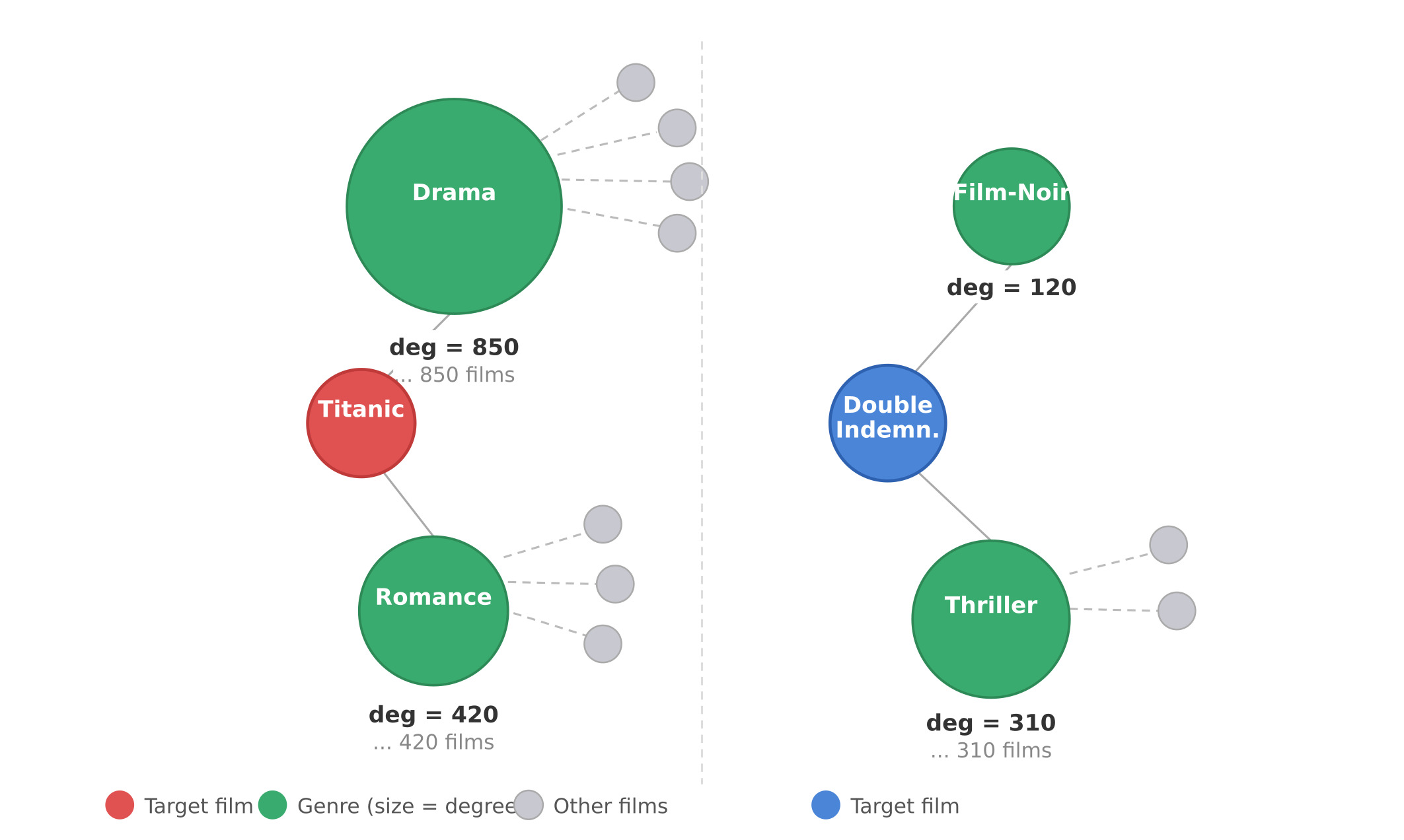}
      \caption{Two films with contrasting genre connectivity. \textit{Titanic} (left) is
      linked to high-degree genres (Drama: 850, Romance: 420), while \textit{Double Indemnity}
      (right) belongs to low-degree genres (Film-Noir: 120, Thriller: 310). Node size
      reflects genre degree.}
      \label{fig:film_examples}
  \end{figure}

 For a head $h$ with tail set $T(h)$, we define its mean tail degree as:
\begin{equation}
    \text{MeanDeg}(h) = \frac{1}{|T(h)|} \sum_{t \in T(h)} \deg(t),
\end{equation}
where $\deg(t)$ is the number of heads connected to tail $t$. A head
linked to high-degree tails sits in a dense neighborhood, providing
the attacker with stronger structural signal.
  Figure~\ref{fig:film_examples} illustrates this contrast with two concrete examples.
  \textit{Titanic} belongs to Drama (deg=850) and Romance (deg=420), giving it a mean genre
  degree of 635. \textit{Double Indemnity} belongs to Film-Noir (deg=120) and Thriller
  (deg=310), giving it a mean genre degree of 215. The size difference in genre nodes directly
  reflects how much structural information is available to the attacker.

\begin{figure}[t]
      \centering
      \includegraphics[width=0.7\textwidth]{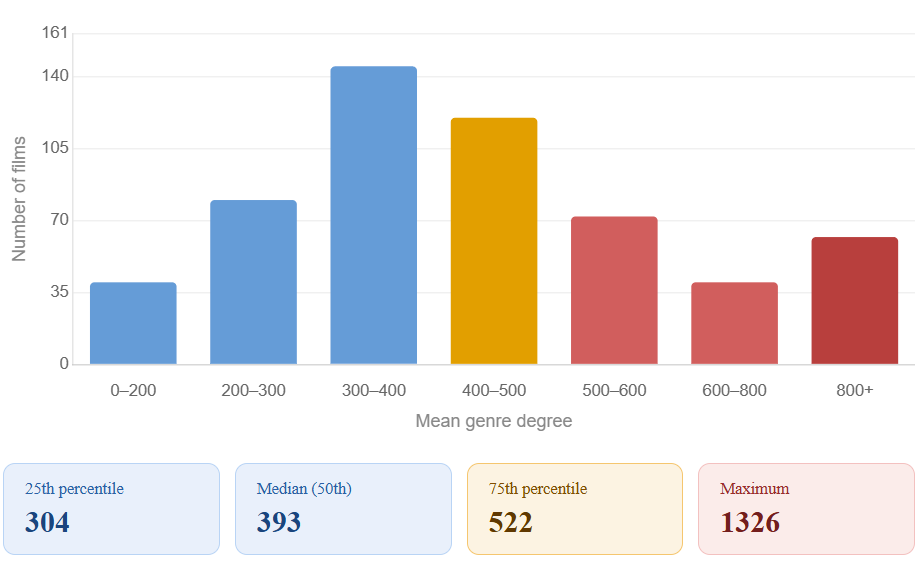}
      \caption{Distribution of mean genre degree across films. The right-skewed distribution
      shows that most films belong to low-to-medium popularity genres, while a small number
      are linked to very popular genres.}
      \label{fig:genre_distribution}
  \end{figure}

  The distribution of $\text{MeanDeg}$ across films is right-skewed (mean 434, median 393,
  25th percentile 304, 75th percentile 522, maximum 1326), confirming that most films have
  low-to-medium genre connectivity, while a minority are linked to very popular genres
  (Figure~\ref{fig:genre_distribution}).

  We partition films into four quartile groups based on $\text{MeanDeg}$ and report Hits@1,
  MRR, and Average Precision (AP) for each group in Table~\ref{tab:unequal_privacy}.

  \begin{table}[t]
  \centering
  \caption{Attack success by mean genre degree quartile.}
  \label{tab:unequal_privacy}
  \begin{tabular}{lcccc}
  \hline
  \textbf{Genre Popularity} & \textbf{Mean Deg.} & \textbf{Hits@1} & \textbf{MRR} & \textbf{AP} \\
  \hline
  Low ($\leq 304$)        & 254 & 43\% & 59\% & 28\% \\
  Below Median (304--393) & 349 & 74\% & 84\% & 36\% \\
  Above Median (393--522) & 457 & 90\% & 93\% & 45\% \\
  High ($> 522$)          & 701 & 92\% & 95\% & 49\% \\
  \hline
  \end{tabular}
  \end{table}

The results show a clear and consistent trend: as genre popularity
increases, so does attack success across all three metrics. Films in
the high popularity group reach 92\% Hits@1 and 95\% MRR, while those
in the low popularity group drop to 43\% Hits@1 and 59\% MRR, showing
that entities in dense regions of the graph are far more exposed than
those in sparse ones.
  Therefore, a uniform edge removal does not provide uniform privacy guarantees. Even after the target relation is fully masked, heads associated with popular tails remain highly vulnerable due to the structural pattern carried by their neighborhood.

%% file: Sections/conclusion.tex
\section{Conclusion \& Future Work} \label{sec:conclusion}

In this paper, we have investigated whether hiding a sensitive relation
from a Knowledge Graph is sufficient to prevent its recovery by a
structural adversary. Our results demonstrate that it is not. Across the two
attack scenarios, structural
features extracted from the public graph, alone, are sufficient to
infer hidden triples with high accuracy. Moreover, our
structural analysis reveals that privacy risk is not uniform:
entities connected to popular, high-degree tail nodes are
substantially more vulnerable than those linked to niche ones,
meaning that uniform edge suppression does not provide uniform
privacy guarantees.

These findings highlight structural leakage as a serious and
underestimated threat to privacy-preserving Knowledge Graph
publication. Our work also raises broader questions about what
constitutes meaningful privacy in graph-structured data, and whether
existing notions of privacy are adequate for the relational complexity
of Knowledge Graphs.While our evaluation is limited to a single dataset, the structural properties driving the attack are generic, and validating these findings on additional KGs from sensitive domains such as biomedical or financial settings remains an important direction for future work. Several defense directions naturally follow from our findings. Graph perturbation and structural anonymization could reduce the discriminative power of the features driving the attack, while differential privacy offers formal guarantees, though its adaptation to heterogeneous KGs remains an open problem. Since privacy risk is non-uniform, any effective defense must be entity-adaptive rather than uniformly applied.

%% file: mybibliography.bib
@inproceedings{bordes2013translating,
author = {Bordes, Antoine and Usunier, Nicolas and Garcia-Dur\'{a}n, Alberto and Weston, Jason and Yakhnenko, Oksana},
title = {Translating embeddings for modeling multi-relational data},
year = {2013},
publisher = {Curran Associates Inc.},
address = {Red Hook, NY, USA},
booktitle = {Proceedings of the 27th International Conference on Neural Information Processing Systems - Volume 2},
pages = {2787–2795},
numpages = {9},
 url = {https://proceedings.neurips.cc/paper_files/paper/2013/file/1cecc7a77928ca8133fa24680a88d2f9-Paper.pdf},
location = {Lake Tahoe, Nevada},
series = {NIPS'13}
}

@inproceedings{toutanova2015observed,
    author    = {Kristina Toutanova and Danqi Chen},
    title     = {Observed versus Latent Features for Knowledge Base and
                 Text Inference},
    booktitle = {Proceedings of the 3rd Workshop on Continuous Vector Space
                 Models and their Compositionality},
    year      = {2015},
    pages     = {57--66}
  }

@inproceedings{wang2019KGAT,
author = {Wang, Xiang and He, Xiangnan and Cao, Yixin and Liu, Meng and Chua, Tat-Seng},
title = {KGAT: Knowledge Graph Attention Network for Recommendation},
year = {2019},
isbn = {9781450362016},
publisher = {Association for Computing Machinery},
address = {New York, NY, USA},
doi = {10.1145/3292500.3330989},
booktitle = {Proceedings of the 25th ACM SIGKDD International Conference on Knowledge Discovery \& Data Mining},
pages = {950–958},
numpages = {9},
location = {Anchorage, AK, USA},
series = {KDD '19}
}

@ARTICLE{9960856,
  author={Lan, Yunshi and He, Gaole and Jiang, Jinhao and Jiang, Jing and Zhao, Wayne Xin and Wen, Ji-Rong},
  journal={IEEE Transactions on Knowledge and Data Engineering}, 
  title={Complex Knowledge Base Question Answering: A Survey}, 
  year={2023},
  volume={35},
  number={11},
  pages={11196-11215},
  doi={10.1109/TKDE.2022.3223858}}

@INPROCEEDINGS{10393039,
  author={Zhou, Yuzhong and Lin, Zhengping and Shi, Jiahao and Yang, Yuliang and Wei, Ronghui},
  booktitle={2023 International Conference on Evolutionary Algorithms and Soft Computing Techniques (EASCT)}, 
  title={Research on the Construction of a Knowledge Graph Information Search Engine in the web Domain}, 
  year={2023},
  volume={},
  number={},
  pages={1-6},
  doi={10.1109/EASCT59475.2023.10393039}}

@inproceedings{backstrom2007wherefore,
author = {Backstrom, Lars and Dwork, Cynthia and Kleinberg, Jon},
title = {Wherefore art thou r3579x? anonymized social networks, hidden patterns, and structural steganography},
year = {2007},
isbn = {9781595936547},
publisher = {Association for Computing Machinery},
address = {New York, NY, USA},
doi = {10.1145/1242572.1242598},
booktitle = {Proceedings of the 16th International Conference on World Wide Web},
pages = {181–190},
numpages = {10},
location = {Banff, Alberta, Canada},
series = {WWW '07}
}

@inproceedings {he2021stealing,
author = {Xinlei He and Jinyuan Jia and Michael Backes and Neil Zhenqiang Gong and Yang Zhang},
title = {Stealing Links from Graph Neural Networks},
booktitle = {30th USENIX Security Symposium (USENIX Security 21)},
year = {2021},
isbn = {978-1-939133-24-3},
pages = {2669--2686},
url = {https://www.usenix.org/conference/usenixsecurity21/presentation/he-xinlei},
publisher = {USENIX Association},
month = aug
}

@misc{wang2021membership,
      title={Membership Inference Attacks on Knowledge Graphs}, 
      author={Yu Wang and Lifu Huang and Philip S. Yu and Lichao Sun},
      year={2022},
      eprint={2104.08273},
      archivePrefix={arXiv},
      primaryClass={cs.AI},
      url={https://arxiv.org/abs/2104.08273}, 
}

@INPROCEEDINGS{xang2019knowledge,
  author={Xiang, Xiayu and Wang, Zhongru and Jia, Yan and Fang, Binxing},
  booktitle={2019 IEEE Fourth International Conference on Data Science in Cyberspace (DSC)}, 
  title={Knowledge Graph-Based Clinical Decision Support System Reasoning: A Survey}, 
  year={2019},
  volume={},
  number={},
  pages={373-380},
  doi={10.1109/DSC.2019.00063}}

@inproceedings{duddu2020quantifying,
author = {Duddu, Vasisht and Boutet, Antoine and Shejwalkar, Virat},
title = {Quantifying Privacy Leakage in Graph Embedding},
year = {2021},
isbn = {9781450388405},
publisher = {Association for Computing Machinery},
address = {New York, NY, USA},
doi = {10.1145/3448891.3448939},
booktitle = {MobiQuitous 2020 - 17th EAI International Conference on Mobile and Ubiquitous Systems: Computing, Networking and Services},
pages = {76–85},
numpages = {10},
location = {Darmstadt, Germany},
series = {MobiQuitous '20}
}

@INPROCEEDINGS{wu2022linkteller,
  author={Wu, Fan and Long, Yunhui and Zhang, Ce and Li, Bo},
  booktitle={2022 IEEE Symposium on Security and Privacy (SP)}, 
  title={LINKTELLER: Recovering Private Edges from Graph Neural Networks via Influence Analysis}, 
  year={2022},
  volume={},
  number={},
  pages={2005-2024},
  doi={10.1109/SP46214.2022.9833806}}

@inproceedings{hu2023quantifying,
author = {Hu, Yuke and Liang, Wei and Wu, Ruofan and Xiao, Kai and Wang, Weiqiang and Li, Xiaochen and Liu, Jinfei and Qin, Zhan},
title = {Quantifying and Defending against Privacy Threats on Federated Knowledge Graph Embedding},
year = {2023},
isbn = {9781450394161},
publisher = {Association for Computing Machinery},
address = {New York, NY, USA},
doi = {10.1145/3543507.3583450},
booktitle = {Proceedings of the ACM Web Conference 2023},
pages = {2306–2317},
numpages = {12},
location = {Austin, TX, USA},
series = {WWW '23}
}

@inproceedings{10.1145/1376616.1376746,
author = {Bollacker, Kurt and Evans, Colin and Paritosh, Praveen and Sturge, Tim and Taylor, Jamie},
title = {Freebase: a collaboratively created graph database for structuring human knowledge},
year = {2008},
isbn = {9781605581026},
publisher = {Association for Computing Machinery},
address = {New York, NY, USA},
doi = {10.1145/1376616.1376746},
booktitle = {Proceedings of the 2008 ACM SIGMOD International Conference on Management of Data},
pages = {1247–1250},
numpages = {4},
location = {Vancouver, Canada},
series = {SIGMOD '08}
}

@ARTICLE{10198395,
  author={Hoang, Anh-Tu and Carminati, Barbara and Ferrari, Elena},
  journal={IEEE Transactions on Dependable and Secure Computing}, 
  title={Protecting Privacy in Knowledge Graphs With Personalized Anonymization}, 
  year={2024},
  volume={21},
  number={4},
  pages={2181-2193},
  doi={10.1109/TDSC.2023.3300360}}

@article{ji2021survey,
  author={Ji, Shaoxiong and Pan, Shirui and Cambria, Erik and Marttinen, Pekka and Yu, Philip S.},
  journal={IEEE Transactions on Neural Networks and Learning Systems}, 
  title={A Survey on Knowledge Graphs: Representation, Acquisition, and Applications}, 
  year={2022},
  volume={33},
  number={2},
  pages={494-514},
  doi={10.1109/TNNLS.2021.3070843}}
